\documentclass[10pt,aps,pre,twocolumn,superscriptaddress]{revtex4-1}
\usepackage[utf8]{inputenc} 
\usepackage{mathtools}
\usepackage{amsfonts}
\usepackage{amsmath}
\usepackage{amsthm}
\usepackage{subfigure}
\usepackage{graphicx}
\usepackage{epsfig}
\usepackage{multirow}
\usepackage{rotating}
\usepackage{amssymb,color}
\newcommand{\IFUFG}{Instituto de F\'{i}sica, Universidade Federal de Goi\'{a}s, Av. Esperan\c{c}a s/n, 74.690-900, Goi\^{a}nia, GO, Brazil}
\newcommand{\IFMT}{Instituto Federal do Mato Grosso, Campus C{\'a}ceres - Prof. Olegário Baldo, Av. dos Ramires s/n, 78200-000, C\'{a}ceres, MT, Brazil}
\newcommand{\DFUFAM}{Departamento de F\'{i}sica, Universidade Federal do Amazonas, 3000, Japiim, 69077-000, Manaus-AM, Brazil}
\newcommand{\IFG}{Instituto Federal de Goi\'as, Rua 76, Centro, Goi\^ania - GO, Brazil}

\begin{document}

\title{Unveiling phase transitions in 1D systems with short-range interactions}

\author{L. S. Ferreira}
\affiliation{\IFUFG}
\author{L. N. Jorge}
\affiliation{\IFMT}
\author{Cl\'{a}udio J. DaSilva}
\affiliation{\IFG}
\author{Minos A. Neto}
\affiliation{\DFUFAM}
\author{A. A. Caparica}
\affiliation{\IFUFG}
\email{caparica@ufg.br}
\date{\today}
\begin{abstract}
The statement that any phase transition is related to the appearance or disappearance of long-range spatial correlations precludes 
a finite transition temperature in one-dimensional (1D) systems. In this paper we demonstrate that the 1D Ising model with short-range exchange interactions exhibits a second-order phase transition at a finite temperature relying on the proper choice of the order parameter. To accomplish this, we combined analytical calculations and high-precision entropic sampling simulations and chose a slightly different order parameter, 
namely the module of the magnetization. Notably, we detected a phase transition with a corresponding critical temperature around 
15 K, which is in excellent agreement with experimental results. Our study indicates that an inappropriate choice of the order 
parameter may mask phase transitions in one-dimensional systems.
\end{abstract}

\maketitle

\section{Introduction}

The impossibility of a phase transition in $(d\leq 2)$-systems with continuous symmetries and short-range interactions has been established 
after the publication of the Mermin-Wagner theorem\cite{Mermin1966}. Particularly, almost all statistical mechanics textbooks, supported by the Perron-Frobenius theorem, prohibits a finite temperature phase transition in one-dimensional (1D) systems with discrete symmetries\cite{Pathria2011}. Nonetheless, many theoretical works have pursuit such transition under different arguments. The development of heteroepitaxy techniques ushered in the manufacturing 
of continuous, extremely low dimensional metal  matrices on non magnetic substrates\cite{khajetoorians2012,Delgado2013,toskovic2016}, allowing experimental 
characterization and theoretical studies of 1D magnetic systems\cite{Vindigni2006}. For instance, the experimental observation 
of an 1D chain of Co atoms deposited on a Pt surface enduring a ferromagnetic phase at a temperature around 15 K\cite{Gambardella2002} 
seems to contradict classical theoretical predictions whether there are true thermodynamic phase transitions in 1D systems with short 
range interactions\cite{Landau1980}. Some attempts to explain this apparent discrepancy have been made. Curilef {\it et al.} 
investigated the critical temperature of the Ising model including a long-range term through a power law that decays over 
large interparticle distances\cite{Curilef2005}. However they made no comparisons to the results of Ref.\cite{Gambardella2002}.
In its turn Li and Liu\cite {Li2006} proposed a Heisenberg model with a large magnetic anisotropy and an additional external magnetic field 
to describe a monatomic chain of Co. Afterwards, Vindigni \textit{et al.}\cite{Vindigni2006} developed a combined experimental 
and theoretical study to explain the behavior of the magnetization of 1D magnetic systems, using the Heisenberg model with an 
anisotropic term, solved by the transfer matrix technique. Recently a study was performed on the magnetic properties of an 
1D Au-Co chain on a copper surface using the density functional theory and kinetic Monte Carlo (KMC)\cite{Kolesnikov2015}.
They performed a KMC simulation of a 1D chain of Heisenberg anisotropic magnets taking the module of the magnetization as the order 
parameter. They found a phase transition at non zero temperature. The authors argued that this does not violate the Mermin-Wagner 
theorem due to the nonzero anisotropic term. 

The foregoing works treated an 1D magnetic chain using a Heisenberg Hamiltonian with an additional anisotropic or a long-range term. 
Notwithstanding the first approximation for an 1D magnetic system should be the Ising model, for which we have now
a new scenario. This model was proposed by Wilhelm Lenz to the doctoral work of his student 
Ernst Ising to describe the behavior of a magnetic material in one dimension. Their solution showed that at zero magnetic field a 
second order phase transition occurs only at $T=0$. However, Lars Onsager solved analytically the 
two-dimensional version of this model and encountered a phase transition at a finite temperature\cite{Onsager1944}. Since then, the 
Ising model became a cornerstone to condensed matter physics and science in general\cite{Noble2015,Aguilera2018,Hoferer2019,Spaide2018,Shin2019}.

In this paper, in contrast with the common understanding, we reveal a phase transition at finite temperature in the 1D Ising model. 
To drive the reasoning, well-established theories should be revisited. In 1966, Kadanoff determined the scaling laws for a second order phase transition of the Ising model\cite{Kadanoff1966} based 
on a Widom's argument\cite{Widom1965} on the consequences of considering the free energy as a homogeneous function of parameters 
describing the proximity to the critical point, such as $\epsilon =(T-T_c)/T_c$. He assumed that a large lattice can be divided 
into smaller ones of still large sizes, but much smaller than the system correlation length. The zeroth order approximation is based on the 
assumption that the free energy of an isolated system of side $L$ is an analytical function for $\epsilon$, 
but singular for the system size. Kadanoff considered two possibilities for the specific heat divergence 
which is given by $\epsilon^{-\alpha}$. The first one assumes the exponent $\alpha \neq 0 $ and the second   
$\alpha=0$, the latter representing the 2D Ising model since the results obtained with $\alpha=0$ corroborate the 
experiments\cite{Kadanoff1966}. Hence, we can state that in a second-order phase transition the specific heat is an 
analytical function of temperature and singularities occur as the system size increases. 

Another important issue is the Landau's theory of second-order phase transitions\cite{Landau1980}, which is based on a 
continuous transition from an ordered state to a disordered one. A second-order phase transition may be related 
to some symmetry property, such as the Curie point where a ferromagnetic substance passes into the paramagnetic phase. 
To describe this event, Landau defined an order parameter as a quantity that assumes nonzero values (positive or negative) 
for an ordered phase and zero for 
the disordered one. Furthermore, he proposed that the singularity is observed when the thermodynamic potential is expanded 
in terms of the order parameter. 

This paper is organized as follows. Section II describes the chose of the order parameter. Section III calculates the partition function using the counting for small lattices and serie analysis. Section IV describes the entropic sampling simulations. Section V combines the analytical and simulational results. Finally, section VI presents our conclusions.

\section{The order parameter}

The Hamiltonian of a one-dimensional Ising model with periodic boundary conditions when an external field is present is such that:
\begin{equation}
\mathcal{H}_L=-J\sum_{i=1}^{L}\sigma_i\sigma_{i+1}-H\sum_{i=1}^{L}\sigma_i,
\end{equation}
where $\sigma_i=\pm1$, $J$ is the coupling constant and $H$ the external magnetic field. Taking the magnetization as the order parameter and $H\neq0$ its value per site in the ground state 
configuration is conditioned to the 
field sign, being $m_+=1$ when the field is positive and $m_-=-1$ when negative.
If $H=0$ there is a singularity where both phases have the same probability. This situation is the same as described by Landau 
for a first order transition, i.e. an equilibrium between two phases. Calculating the mean magnetization  at $T=0$ we have
\begin{equation}\label{eq.1}
\langle m \rangle=m_{+}P_{\{ \uparrow \uparrow \uparrow \}}+m_{-}P_{\{ \downarrow \downarrow \downarrow \}},
\end{equation}
where $P_{\{ \uparrow \uparrow \uparrow \}}$ ($P_{\{ \downarrow \downarrow \downarrow \}}$) denotes the probability of finding a configuration 
with all spins up (down). Substituting the values of  $m_+$ and $m_-$ we obtain $\langle m \rangle=0$. 
At high temperatures the most probable configurations are those where the energies match $k_BT$ and in the limit
$T \rightarrow \infty$ they are configurations where $\mathcal{H}_L=0$, the maximum energy. 
For these configurations the magnetization is zero since the number of positive and negative spins is the same. 
Hence, if $H=0$, the magnetization does not behave as an order parameter in the study of a second order phase transition
in the 1D Ising model, since it does not distinguish an ordered configuration from a disordered one.

Since the magnetization cannot be used as the order parameter for $H=0$ the corresponding correlation 
length is not a proper criterion for validate the existence of a phase transition. To understand the reason, let us 
take a look at the relation between the correlation length and the spin-spin correlation function.
   
The spin-spin correlation function is defined, for lattice systems, as
\begin{equation}\label{cdc}
g(|\vec{r}_i-\vec{r}_j|)= \langle \sigma_i\sigma_j \rangle - \langle \sigma_i \rangle \langle \sigma_j \rangle.
\end{equation}
Hence, for a configuration in the ordered phase, $g(\vec{r})$ must be zero. Let us exam its value for the 1D Ising 
model when $H=0$ and $T=0$. The term $\langle \sigma_i \rangle=\langle \sigma_j \rangle=m$ is null due to the 
existence of two antisymmetric ordered configurations in the ground state, matching the Eq. \ref{eq.1}. This 
statement is based on the fact that the probabilities of finding the system in an ordered configuration with spins 
up or down are equal. Clearly, both configurations are not present simultaneously in real systems. However, in the 
Transfer Matrix calculation one considers the superposition of both configurations, yielding a zero average total 
magnetization. This result is unrealistic since in the thermodynamic limit the system phase is determined by the 
initial conditions and only one ordered configuration is allowed. Several textbooks showing the solution for the 1D 
Ising model consider its experimental realization where two antisymmetric solutions are possible, allowing the 
correlation function to be zero at $T=0$ \cite{Baxter1982,Goldenfeld1992,Christensen2005,Pathria2011}. However, they 
use the Transfer Matrix result for $T>0$.

The bottom line is simply whether or not $\langle \sigma_i \rangle$ is zero for all $T$. Textbooks show that the average magnetization is zero for any temperature $T\neq 0$ and, using a mathematical trick,  promptly prove that it assumes the value $+1$ or $-1$ for $T = 0$ \cite{Goldenfeld1992,Christensen2005,Pathria2011}. This clearly suggest that for an extremely low temperature, say $T = 10^{-23} \text{ K}$, the magnetization is null, however for $T = 0$ it will be either $+1$ or $-1$. This is certainly a non-physical situation. The entire scientific community has accepted this assertion through decades. Hence,  it will be very difficult to convince anyone that this is an inconsistent strategy. Definitely, this mathematical maneuver is tailor-made to conclude that $\langle \sigma_i \rangle\langle \sigma_j \rangle=1$, allowing to achieve $T_c=0$.

In its turn, the term $\langle \sigma_i\sigma_j \rangle$ is always positive. Thus, $g(\vec{r})=\langle \sigma_i
\sigma_j \rangle$ is always greater than zero for the Ising model when $T=0$ contravening the requirement of 
$g(\vec{r})=0$ for an ordered configuration. This result is acceptable when we use the results from the Transfer 
Matrix for any value of $T$.

The correlation function is, therefore, not precisely defined for $H=0$ and its value, obtained from the Transfer 
Matrix, seems to be not appropriate. Another important aspect is that the definition of the correlation length is 
similar to the magnetic susceptibility, hence if the latter diverges, the former also diverges. It is related to 
the magnetic susceptibility as
\begin{equation}
\xi=k_BT\chi,
\end{equation}
where $\chi=\lim_{H\rightarrow 0}(\frac{\partial M}{\partial H})$. 

Otherwise, if we choose as the order parameter the module of the magnetization per site 
\begin{equation}
|m|=\frac{1}{L}|\sum_{i=1}^{L}\sigma_i|,
\end{equation}
such that  $|m_+|=|m_-|=1$, we will obtain at $T=0$ $\langle |m| \rangle=
P_{\{ \uparrow \uparrow \uparrow \}}+P_{\{ \downarrow \downarrow \downarrow \}}=1$, whereas for an disordered configuration we get
$\langle |m| \rangle=0$. Thus, the order parameter that indeed distinguishes an ordered phase from a disordered one is the module of the magnetization. This is a well-known fact for the 2D case and was used recently to explore a phase transition in a 1D Heinsenberg chain of spins\cite{Kolesnikov2015}.

\section{Partition function}

If we adopt the module of the magnetization as the order parameter and write the Hamiltonian in terms of this order parameter
\begin{equation}
\mathcal{\tilde{H}}=-J\sum_{i=1}^{L}\sigma_i\sigma_{i+1}-H|\sum_{i=1}^{L}\sigma_i|,
\label{hamil}
\end{equation}
then we can calculate a partition function $\mathcal{\tilde{Z}}$ using $\mathcal{\tilde{H}}$ such
that when $H=0$ it will yield reliable results. 
In this case the solution via transfer matrix becomes unfeasible, because the module in the second term 
is not factorable. Since an exact solution for an infinite lattice in this case is not available, what 
can be done is the direct calculation of the partition function for small sizes.

Defining $E_j$ and $M_j$ as
\begin{equation}
 E_j=\sum_{i=1}^{L} \sigma_i \sigma_{i+1}~~\text{and}~~
 M_j=\sum_{i=1}^{L} \sigma_i
\end{equation}
where the index $j$ denotes a particular configuration, we can write $\mathcal{\tilde{H}}=-JE_j-H|M_j|$
and the partition function may be written as
\begin{equation}
\tilde{Z}=\sum_j e^{\beta(JE_j+H|M_j|)},
\end{equation} 
where the sum runs over all possible configurations. In terms of the joint density of states $g_L(E,M)$ for a finite lattice
of size $L$
\begin{align}
 \tilde{Z}_L=&\sum_{E=-L}^{L} \sum_{M=-L}^{L} g_L(E,M) e^{\beta(JE+H|M|)}\notag \\
          =&\sum_{E=-L}^{L} \sum_{M=0}^{L} \tilde{g}_L(E,M) e^{\beta(JE+HM)},
\end{align}
where
\[
   \tilde{g}_L(E,M)=
\begin{cases}
    g_L(E,-M)+g_L(E,M),& \text{if } M\neq 0;\\
    g_L(E,M),& \text{if } M = 0.
\end{cases}
\]
\begin{figure}[t]
 \includegraphics[scale=0.45, angle=0]{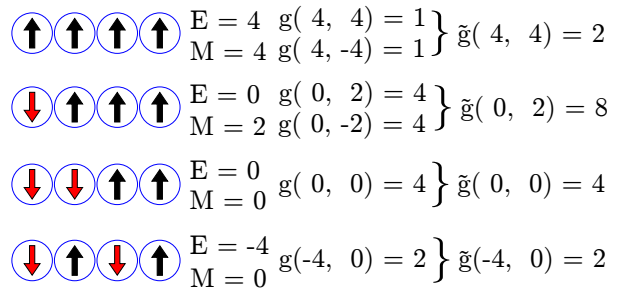}
\caption{Spin configurations for $L=4$. The first represents two configurations, the second eight, and so forth. 
$g(E,M)$ enumerates the states considering the magnetization as the order parameter, while $\tilde g(E,M)$ 
represents the degeneracy when the order parameter is the module of the magnetization.\label{fig0}}
\end{figure}

In Fig. \ref{fig0} we depict the possible configurations for a $L=4$ lattice, from which we can write the partition function for this 
lattice size as
\begin{flalign*}
\tilde{Z}_4=&\tilde{g}_4(4,4)e^{4\beta H}e^{4\beta J}+ \tilde{g}_4(0,2)e^{2\beta H}\\ 
&+\tilde{g}_4(0,0)+\tilde{g}_4(-4,0)e^{-4\beta J}\\
=&2e^{4\beta H}e^{4\beta J}+ 8e^{2\beta H}+4+2e^{-4\beta J},
\end{flalign*}
where $\tilde{g}_4(E,M)$ is the number of configurations with dimensionless interaction energy $E$ and order parameter $M$. 
One can see that $\tilde{g}_4(4,4)=2$, $\tilde{g}_4(0,2)=8$, $\tilde{g}_4(0,0)=4$ and $\tilde{g}_4(-4,0)=2$, where the first term corresponds to 
the ordered states, the second one to the configurations with one unpaired spin and the remaining to configurations with two 
unpaired spins, in which case they may be neighbors or not, corresponding to the terms $\tilde{g}_4(0,0)$ and $\tilde{g}_4(-4,0)$, 
respectively.

Using the same counting process for $L=8$ we obtain
\begin{flalign}
\tilde{Z}_8=&2e^{8\beta H}e^{8\beta J}+ 16e^{6\beta H}e^{4\beta J}+e^{4\beta H}(16e^{4\beta J}+40)\notag\\ 
&+e^{2\beta H}(16e^{4\beta J}+64+32e^{-4\beta J})\notag\\
&+8e^{4\beta J}+36+24e^{-4\beta J}+2e^{-8\beta J}.
\end{flalign}

\section{Simulational Details}

Entropic sampling simulations are based on the Wang-Landau method\cite{Wang2001} where we introduced some changes to improve accuracy and help 
saving CPU time\cite{Caparica2012,Caparica2014,Ferreira2018}. We halt the simulations when the sixteenth Wang-Landau level $(f_{15})$ 
becomes flat. Once the joint density of states is available one can obtain the canonical averages as
\begin{flalign}\label{canonicalmean1}
 \langle E^{k}M^{l} \rangle_{T,H}=&\dfrac{\sum_{E,M=-L}^{L}E^{k}M^{l}g(E,M)e^{-\beta(-JE-HM)}}{\sum_{E,M=-L}^{L}g(E,M)e^{-\beta(-JE-HM)}},\\
 & k,l=0,1,2,..., \notag
\end{flalign}
which are present in the estimate of thermodynamic quantities as mean energy, magnetization, heat capacity, susceptibility and cumulants. 

The joint density of states for a lattice of size $L$ can be used in Eq.\eqref{canonicalmean1}. Also, setting $k=0$, $l=1$ 
and taking $|M|$ before $g_L(E,M)$, we obtain via simulations the same magnetizations for these lattice sizes for $H=0$ as
\begin{flalign}\label{canonicalmean2}
 \langle |M| \rangle_{T,L}=&\dfrac{\sum_{E=-L}^{L} \sum_{M=-L}^{L}|M|g_L(E,M)e^{\beta JE}}{\sum_{E=-L}^{L} \sum_{M=-L}^{L}g_L(E,M)e^{\beta JE}}.
\end{flalign}

\section{Results}

In the present work we restrict ourselves to magnetic spin models, nonetheless the proper choice of the order parameter may trigger important features in a variety of systems. This was evinced recently with the observation of the coexistence of first- and second-order phase transition in a correlated oxide\cite{Post2018}. Another challenge is the study of the Bell-Lavis lattice model to water where the choice of the order parameter is crucial to unveil the metastable behavior of the system\cite{Simenas2014,Palmer2018}. It is worth noting that the Perron-Frobenius theorem does not apply to this case, since its 
demonstration assumes that the partition function can be factored.

Any thermodynamical quantity can be obtained once one achieves the partition function. Here we have computed the average value of the module of the magnetization $\tilde M_L$, since this is the appropriate quantity to observe a phase transition in our model. We can easily get $\tilde M_L$ using the Helmholtz free energy $\tilde F_L=-k_BT\ln \tilde Z_L$, with 
$\tilde M_L=\lim_{H\rightarrow 0}(-\frac{\partial \tilde F_L}{\partial H})$, yielding
\begin{equation}
 \tilde M_4 =4~\frac{1+2e^{-4\beta J}}{1+6e^{-4\beta J}+e^{-8\beta J}}
 \label{m4}
\end{equation}
and 
\begin{equation}
 \tilde M_8=8~\frac{1+12e^{-4\beta J}+18e^{-8\beta J}+4e^{-12\beta J}}{1+28e^{-4\beta J}+70e^{-8\beta J}+28e^{-12\beta J}+e^{-16\beta J}}.
 \label{m8}
\end{equation}
\begin{figure}[t]
 \includegraphics[scale=0.6, angle=-90]{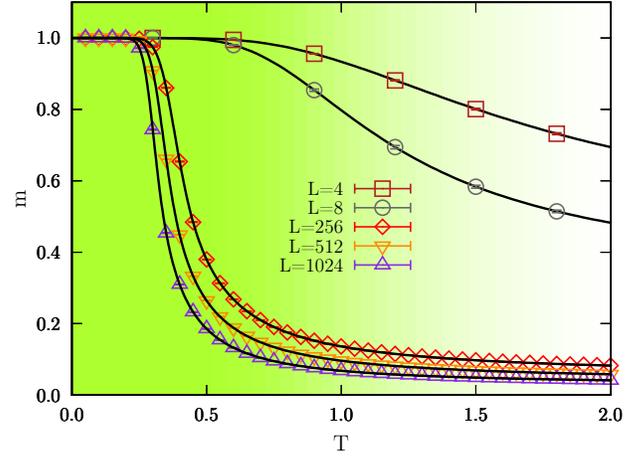}
\caption{Magnetization as function of temperature exhibiting a clear finite-size scaling behavior. 
The lines represent the analytical results, and the dots the simulated ones. The error bars are smaller than the symbols.\label{fig1}}
\end{figure}

Inspecting the numerators and denominators of the two quantities above we see that they can be described by the following series
\begin{align}
  N &= \sum_{k=0}^{L/2}\binom{\lfloor\frac{L-1}{2}\rfloor}{k}\binom{L/2}{k}e^{-4k\beta J} \\
  D &= \sum_{k=0}^{L/2}\binom{L}{2k}e^{-4k\beta J},
\end{align}
where $\lfloor \rfloor$ denotes the largest integer not greater than the argument and $L$ should be a power of $2$. 

In order to confirm these predictions, we carried out entropic sampling simulations for $L=4, ~8, ~256, ~512$, and $~1024$ to construct the joint density of states. In Fig.\ref{fig1} we display the analytical results and those of the simulations for these lattice sizes, assuming $J=1$. One can see that they match with great accuracy. It is also clear that the system displays finite-size effects, suggesting a finite-size scaling study. It is important to note that for lattice sizes greater than $8$ the simulations corroborate our conjectures on the series pointed out above. The analytical expression of the partition function for $L>8$ and not a power of $2$ is unattainable. Nonetheless, the simulations are free to go ahead to as large and diverse lattices as necessary.

We can also obtain a susceptibility of the order parameter as $\tilde{\chi_L}=\lim_{H\rightarrow 0}(\frac{\partial \tilde M_L}{\partial H})$ yielding
\begin{equation}
\tilde{\chi_4}=16\beta\left(\frac{1 + e^{-4J\beta }}{1 + 6 e^{-4J\beta } + e^{-8J\beta }} - \tilde{M}_4^2\right),
\end{equation}
and
\begin{align}
&\tilde{\chi_8}= \notag \\ 
& 64\beta\left(\frac{1 + 7 e^{-4J\beta} + 7 e^{-8J\beta} + e^{-12J\beta}}{1 + 28 e^{-4J\beta} + 70 e^{-8J\beta} + 28 e^{-12J\beta} + e^{-16J\beta}} - \tilde{M}_8^2\right).
\end{align}
Notice that the second term is exactly the square of the average magnetization. Then, the first term should be $\langle M^2_L \rangle$. The denominator is equal to the magnetization and the numerator can be written as
\begin{equation}
N_{\chi}=\sum_{k=0}^{L/2}\binom{L}{2k+1}\frac{1}{L}.
\end{equation}
These series for the numerator and denominator can be found in Refs. \cite{Sloane2011,Hanna2006,Renner2017}.

\begin{figure}[!t]
\centering
\includegraphics[scale=0.6,angle=-90]{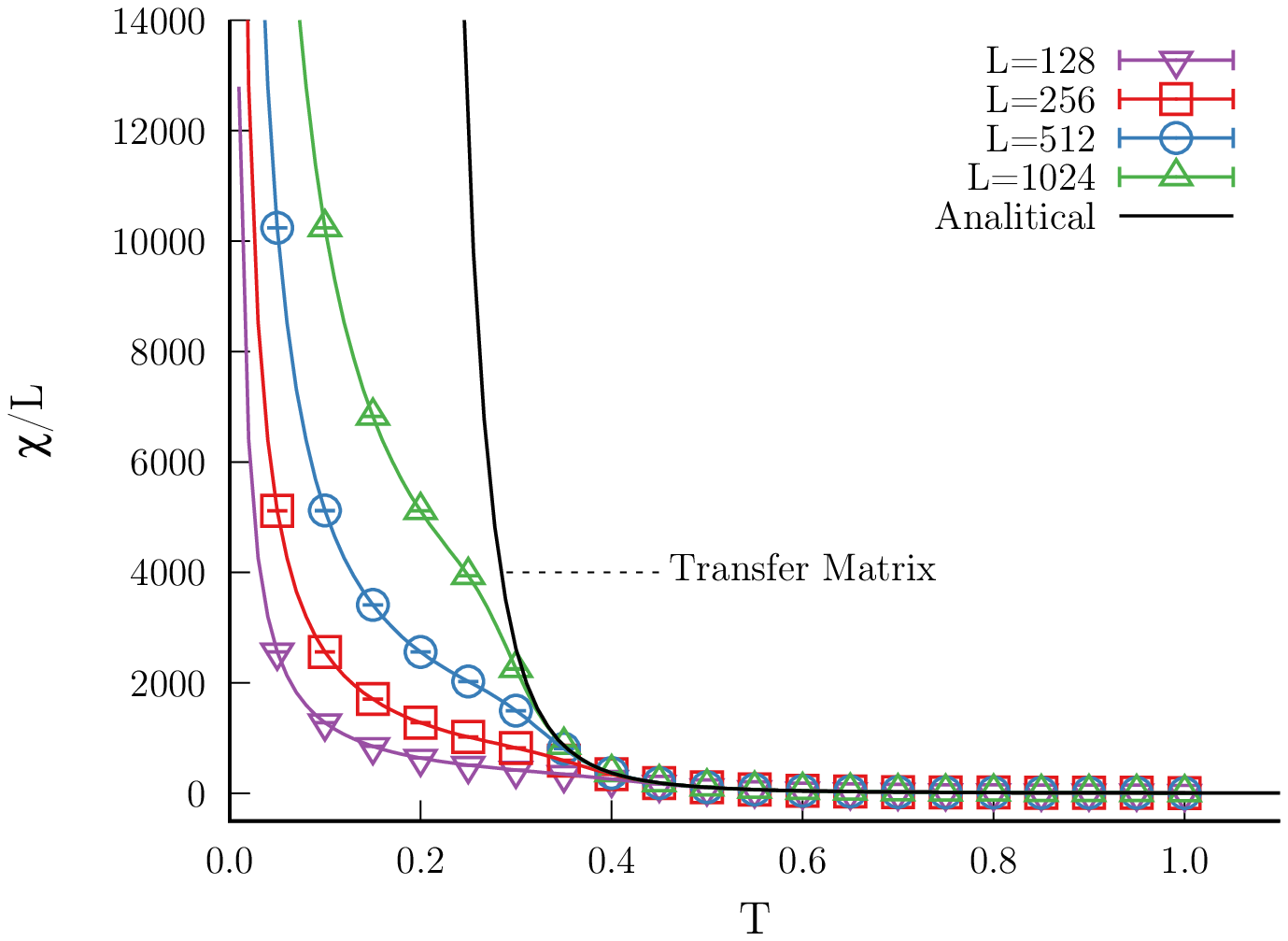}
\includegraphics[scale=0.6,angle=-90]{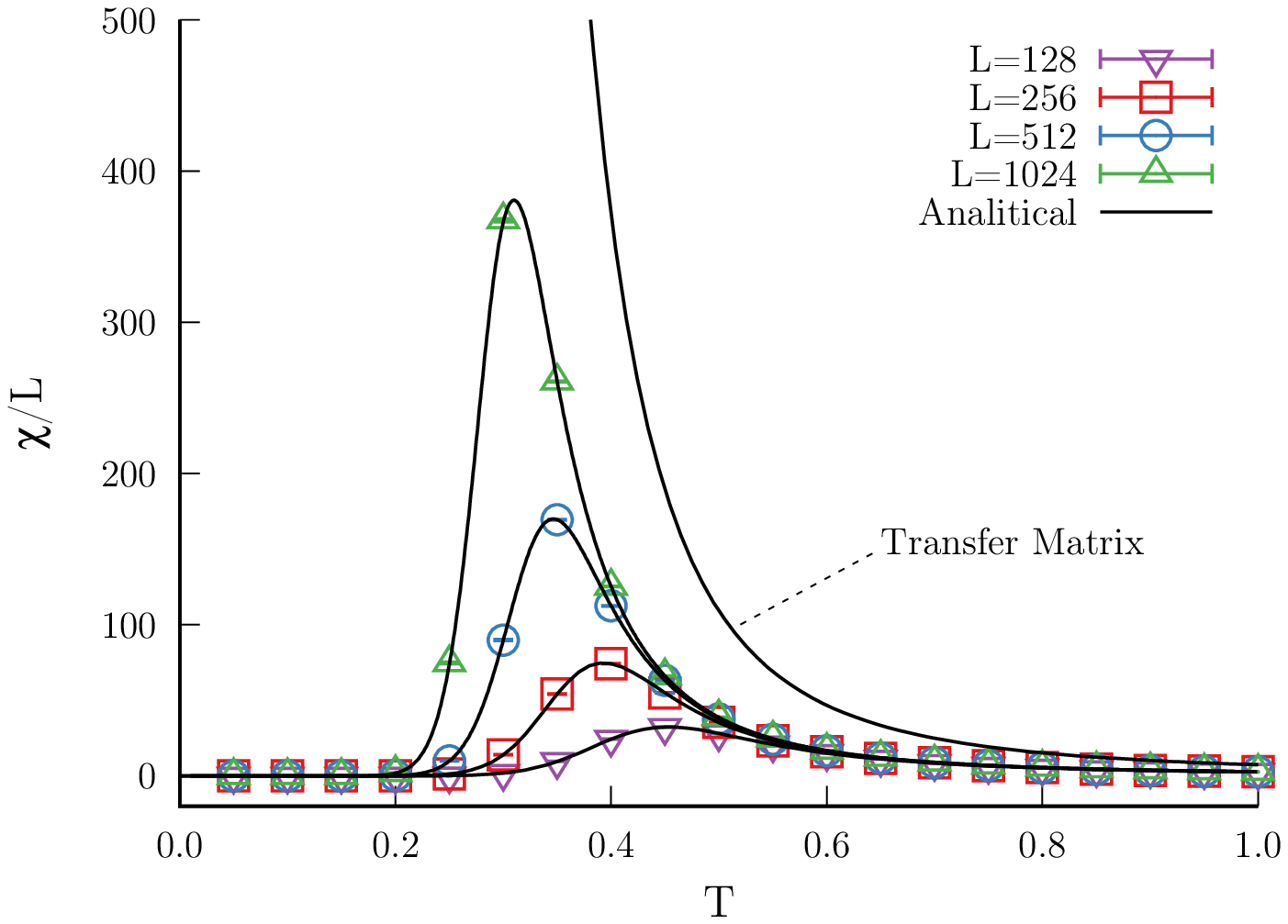}
\caption{Magnetic susceptibility obtained from both simulation and Transfer Matrix technique. (up) Using the magnetization as the order parameter and (bottom) using $m'=|\sum_i \sigma_i|$ as the order parameter. Here, the lines represent the susceptibility obtained using the series and the symbols the simulation results.\label{fig2}}
\end{figure}

In Fig. \ref{fig2}, we show the behavior of the magnetic susceptibility considering both the standard and the new order parameter $m'=|\sum_i \sigma_i|$ obtained from simulation and the Transfer Matrix technique for the 1D Ising model.

One can see that the simulational results using the magnetization as the order parameter corroborate the Transfer Matrix ones, although an FSS behavior is noticeable.  Also, the susceptibility shows a peak around the region where the analytical value diverges when we use the absolute value of the magnetization as the order parameter. It is important to state that both results were taken from the same simulational procedure, the only difference being the order parameter used.
\begin{figure}[t]
\centering
\includegraphics[scale=0.6,angle=-90]{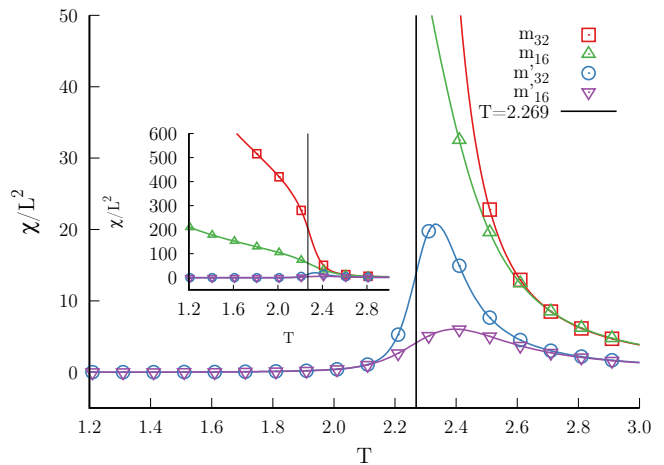}
\caption{Contrast between magnetic susceptibility simulated using different order parameters, namely the magnetization and $m'=|\sum_i \sigma_i|$ for the 2D Ising model with $L=16$ and $L=32$.\label{fig3}}
\end{figure}

The same behavior appears in the 2D Ising model, as one can see in Fig. \ref{fig3}, where is shown the magnetic susceptibility for square lattices of sizes $L=16$ and $32$. One can see that the divergence of the susceptibility starts near the critical temperature, even though its divergence continues for $T<T_c$, up to $T=0$. This can be considered as an FSS effect similar to the 1D case. Notice also that without a maximum in the susceptibility, to determine the critical temperature is unfeasible. Choosing the order parameter as $m'=|\sum_i \sigma_i|$ the divergence of the susceptibility is associated with the limit $L\rightarrow \infty$ as suggested by Kadanoff\cite{Kadanoff1966}.

Since the simulation and analytical results agree perfectly, we will use the simulation procedure to study the finite-size behavior. Following Refs.\cite{Chen1993,Caparica2000a} we can define a set of thermodynamic quantities related to logarithmic 
derivatives of the magnetization that scales as 
\begin{equation}\label{mn}
V_j\approx \frac{1}{\nu}\ln L+\mathcal{V}_j(tL^{1/\nu})
\end{equation}
for $j=1,2,...,6$, where $t=(T-T_c)/T_c$ is the reduced temperature.
These cumulants allow to estimate the critical exponent $\nu$ before determining the critical temperature. 
\begin{figure}[t]
 \includegraphics[scale=0.6, angle=-90]{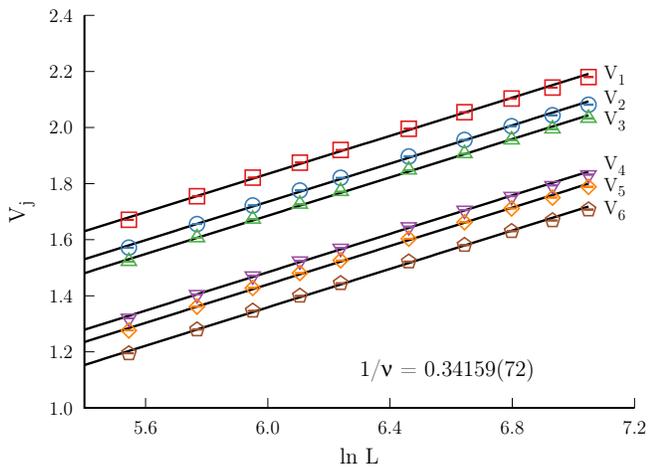}
\caption{Size dependence of the maxima of $V_j$. The slopes yield $1/\nu$. The error bars are smaller than the symbols.\label{fig4}}
\end{figure}
\begin{figure}[!t]
 \includegraphics[scale=0.6, angle=-90]{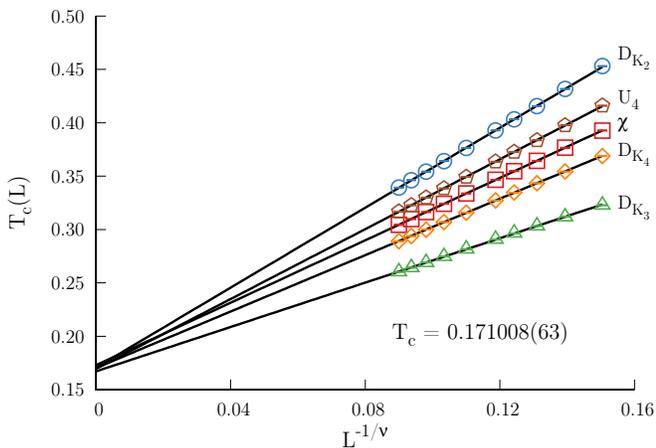}
\caption{Size dependence of the locations of the extrema in different thermodynamic quantities with $\nu=2.92749$.
The error bars are smaller than the symbols.\label{fig5}}
\end{figure} 

We performed simulations for $L= 256,~320,~384,$ $~448,~512,~640,~768,~896,~1024$, and $1152$ with
$n=24 ,~24 ,~20 ,~20 ,~20 ,~16 ,~16 ,~16 ,~12$, and $12$ independent runs for each size, respectively.
In Fig. \ref{fig4} we depict the behavior of these quantities, obtaining $1/\nu=0.34159(72)$ as the mean value of the slopes of the six best
fitting straight lines, yielding $\nu=1/(1/\nu)\pm1/(1/\nu)^2 \Delta (1/\nu)=2.9274(33)$. The exponent $\nu$ was obtained in studies of  
1D Ising model with long-range interactions, as $J_{ij}=J_0/|i-j|^{1+\sigma}$, such that the larger is $\sigma$ the closer the system is
of the short-range interactions. In Ref.\cite{Glumac1989} they obtained $\nu=2.602721$ for $10$ interacting spins and $\sigma=1.0$, while in Ref.\cite{Bayong1999}
the estimate using Monte Carlo histogram method and $\sigma=1.0$ is $\nu=2.42(1)$. These results suggest that our $\nu$ consists in a short-range
limit for the long-range interactions systems.

Now, with the correlation length critical exponent $\nu$ well determined we can use the finite-size scaling relation
\begin{equation}\label{tc}
T_c(L)=T_c+aL^{1/\nu},
\end{equation}
where $a$ is a constant, to determine $T_c$ for an infinite lattice.
This relation is valid for the quantities $\chi$, $D_{K_2}$, $D_{K_3}$, $D_{K_4}$, and $U_4$, defined in
Refs.\cite{Chen1993,Caparica2000a}. Theses results are shown in the Fig. \ref{fig5}. For an infinite lattice ($L\rightarrow \infty$) we obtain five estimates
for the critical temperature, the mean value yielding $T_c = 0.171008(63)$. This finite value for the critical temperature contradicts the Transfer Matrix result. There is only one way to support this result is to find a one-dimensional magnetic material that bears a phase transition at any finite temperature. 

The findings of a finite-temperature phase transition for 1D systems with short-range interactions are not by any means a surprise in the scientific community, as can be seen in the investigation of domain walls in Fe nanostripes\cite{Shen1997} or the study of a ferromagnetic order in a line of cobalt atoms\cite{Gambardella2002}. Thereon, several theoretical works tried to explain the phenomenum of ferromagnetism in one-dimension using the Heisenberg spin to model the systems\cite{ Li2006,Vindigni2006,Kolesnikov2015}.

An excellent candidate to design a one-dimensional Ising chain is cobalt due to the strong anisotropic field\cite{Wolf2000}. A study about the compound Ca$_3$Co$_2$O$_6$, using orbital image technique, showed that the high anisotropic field freezes the spin in the easy axis direction\cite{Leedahl2019}. This behavior endorses the comparison between cobalt chains and the 1D Ising model, causing the use of the Heisenberg model not strictly necessary. Therefore, we must look more closely at Gambardella's \cite{Gambardella2002} work.

To investigate the cobalt chains Gambardella et al. used the X-ray magnetic circular dichroism (XMCD) technique and found that the magnetic spin moment is $m_s\sim 2.03 \mu_B$ and the magnetic anisotropic energy per atom is $2.0 \pm 0.2 ~meV$, which is an extremely high value when compared to the bulk one. These values confirm the results obtained in Ref. \cite{Leedahl2019}. Gambardella et al. also obtained a blocking temperature of $T_B=15\pm 5$ K, for which the ferromagnetic order disappear. Therefore, we will use $2J=15$ meV to compare our results with the one obtained by Gambardella et al. We then obtained our critical temperature as $ T_c = 0.171008 (63) J/k_B=14.8835(55)$ K, where 
$k_B=8.6173324(78)\times10^{- 5}$ eV/K, in excellent agreement with the experiment.
\begin{figure}[!t]
 \includegraphics[scale=0.6, angle=-90]{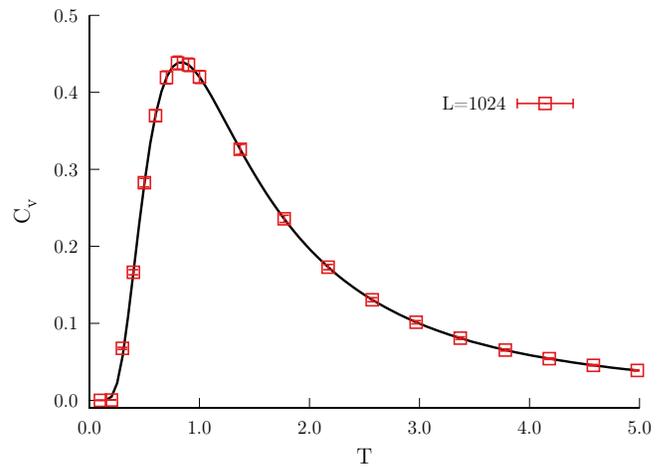}
\caption{Specific heat as a function of temperature for $L=1024$. The solid line represents the exact solution by
transfer matrix technique.\label{fig6}}
\end{figure}

It is worth noting that in this work we have analyzed only the discontinuity on the order parameter since quantities such as specific heat show no finite size behavior, in contrast with 2D systems. In Fig. \ref{fig6} we show how the simulation result for $L=1024$ lattice fits that obtained analytically by transfer matrix technique. This behavior is known as Schottky effect, first observed in two-state systems, and in general is not accepted as a phase transition signature. More complex systems also exhibit this effect and a complete understanding of its origin is still lacking\cite{Simenas2015}. We believe that the appearance of this effect in our model is due to the fact that the first term in Eq. \ref{hamil} is still factorable when computing the partition function, what does not happen in higher dimensions.

\section{Conclusions}
 
In summary, we see that the choice of the order parameter for the 1D Ising model as the module of the
magnetization and not simply as the magnetization, whose average does not distinguish the ordered state from 
the disordered one, reveals the existence of a second-order phase transition with a finite and well-defined critical 
temperature, in contrast to the belief that such a transition will never be exhibited, 
as stated by almost all statistical mechanics textbooks and supported by the Perron-Frobenius theorem. However, as explained above, we see that it does not apply to the present case, 
once the factorization of the partition function becomes impossible when we take the module of the magnetization as 
the order parameter instead of the magnetization itself.

Our results open up a wide range of opportunities to understand phase transitions in one dimension and 
may have potential ramifications and applications, leading subsequently to possible advances in a much broader context. 

\section*{Acknowledgement}

We thank J.N. Teixeira Rabelo, M.S. Carrião, and Z. Glumac for helpful comments and suggestions. We acknowledge the computer resources
provided by LCC-UFG. L.S.F. acknowledges the support by CAPES and M.A.N. by FAPEAM and CNPq.
 

\begin{thebibliography}{40}%
\makeatletter
\providecommand \@ifxundefined [1]{%
 \@ifx{#1\undefined}
}%
\providecommand \@ifnum [1]{%
 \ifnum #1\expandafter \@firstoftwo
 \else \expandafter \@secondoftwo
 \fi
}%
\providecommand \@ifx [1]{%
 \ifx #1\expandafter \@firstoftwo
 \else \expandafter \@secondoftwo
 \fi
}%
\providecommand \natexlab [1]{#1}%
\providecommand \enquote  [1]{``#1''}%
\providecommand \bibnamefont  [1]{#1}%
\providecommand \bibfnamefont [1]{#1}%
\providecommand \citenamefont [1]{#1}%
\providecommand \href@noop [0]{\@secondoftwo}%
\providecommand \href [0]{\begingroup \@sanitize@url \@href}%
\providecommand \@href[1]{\@@startlink{#1}\@@href}%
\providecommand \@@href[1]{\endgroup#1\@@endlink}%
\providecommand \@sanitize@url [0]{\catcode `\\12\catcode `\$12\catcode
  `\&12\catcode `\#12\catcode `\^12\catcode `\_12\catcode `\%12\relax}%
\providecommand \@@startlink[1]{}%
\providecommand \@@endlink[0]{}%
\providecommand \url  [0]{\begingroup\@sanitize@url \@url }%
\providecommand \@url [1]{\endgroup\@href {#1}{\urlprefix }}%
\providecommand \urlprefix  [0]{URL }%
\providecommand \Eprint [0]{\href }%
\providecommand \doibase [0]{http://dx.doi.org/}%
\providecommand \selectlanguage [0]{\@gobble}%
\providecommand \bibinfo  [0]{\@secondoftwo}%
\providecommand \bibfield  [0]{\@secondoftwo}%
\providecommand \translation [1]{[#1]}%
\providecommand \BibitemOpen [0]{}%
\providecommand \bibitemStop [0]{}%
\providecommand \bibitemNoStop [0]{.\EOS\space}%
\providecommand \EOS [0]{\spacefactor3000\relax}%
\providecommand \BibitemShut  [1]{\csname bibitem#1\endcsname}%
\let\auto@bib@innerbib\@empty
\bibitem [{\citenamefont {Mermin}\ and\ \citenamefont
  {Wagner}(1966)}]{Mermin1966}%
  \BibitemOpen
  \bibfield  {author} {\bibinfo {author} {\bibfnamefont {N.~D.}\ \bibnamefont
  {Mermin}}\ and\ \bibinfo {author} {\bibfnamefont {H.}~\bibnamefont
  {Wagner}},\ }\href {\doibase 10.1103/PhysRevLett.17.1307} {\bibfield
  {journal} {\bibinfo  {journal} {Phys. Rev. Lett.}\ }\textbf {\bibinfo
  {volume} {17}},\ \bibinfo {pages} {1307} (\bibinfo {year}
  {1966})}\BibitemShut {NoStop}%
\bibitem [{\citenamefont {Pathria}\ and\ \citenamefont
  {Beale}(2011)}]{Pathria2011}%
  \BibitemOpen
  \bibfield  {author} {\bibinfo {author} {\bibfnamefont {R.~K.}\ \bibnamefont
  {Pathria}}\ and\ \bibinfo {author} {\bibfnamefont {P.~D.}\ \bibnamefont
  {Beale}},\ }\href@noop {} {\emph {\bibinfo {title} {Statistical mechanics}}}\
  (\bibinfo  {publisher} {Elsevier},\ \bibinfo {year} {2011})\BibitemShut
  {NoStop}%
\bibitem [{\citenamefont {Khajetoorians}\ \emph {et~al.}(2012)\citenamefont
  {Khajetoorians}, \citenamefont {Wiebe}, \citenamefont {Chilian},
  \citenamefont {Lounis}, \citenamefont {Blügel},\ and\ \citenamefont
  {Wiesendanger}}]{khajetoorians2012}%
  \BibitemOpen
  \bibfield  {author} {\bibinfo {author} {\bibfnamefont {A.~A.}\ \bibnamefont
  {Khajetoorians}}, \bibinfo {author} {\bibfnamefont {J.}~\bibnamefont
  {Wiebe}}, \bibinfo {author} {\bibfnamefont {B.}~\bibnamefont {Chilian}},
  \bibinfo {author} {\bibfnamefont {S.}~\bibnamefont {Lounis}}, \bibinfo
  {author} {\bibfnamefont {S.}~\bibnamefont {Blügel}}, \ and\ \bibinfo
  {author} {\bibfnamefont {R.}~\bibnamefont {Wiesendanger}},\ }\href@noop {}
  {\bibfield  {journal} {\bibinfo  {journal} {Nature Physics}\ } (\bibinfo
  {year} {2012})}\BibitemShut {NoStop}%
\bibitem [{\citenamefont {Delgado}\ \emph {et~al.}(2013)\citenamefont
  {Delgado}, \citenamefont {Batista},\ and\ \citenamefont
  {Fern\'andez-Rossier}}]{Delgado2013}%
  \BibitemOpen
  \bibfield  {author} {\bibinfo {author} {\bibfnamefont {F.}~\bibnamefont
  {Delgado}}, \bibinfo {author} {\bibfnamefont {C.~D.}\ \bibnamefont
  {Batista}}, \ and\ \bibinfo {author} {\bibfnamefont {J.}~\bibnamefont
  {Fern\'andez-Rossier}},\ }\href {\doibase 10.1103/PhysRevLett.111.167201}
  {\bibfield  {journal} {\bibinfo  {journal} {Phys. Rev. Lett.}\ }\textbf
  {\bibinfo {volume} {111}},\ \bibinfo {pages} {167201} (\bibinfo {year}
  {2013})}\BibitemShut {NoStop}%
\bibitem [{\citenamefont {Toskovic}\ \emph {et~al.}(2016)\citenamefont
  {Toskovic}, \citenamefont {van~den Berg}, \citenamefont {Spinelli},
  \citenamefont {Eliens}, \citenamefont {van~den Toorn}, \citenamefont
  {Bryant}, \citenamefont {Caux},\ and\ \citenamefont {Otte}}]{toskovic2016}%
  \BibitemOpen
  \bibfield  {author} {\bibinfo {author} {\bibfnamefont {R.}~\bibnamefont
  {Toskovic}}, \bibinfo {author} {\bibfnamefont {R.}~\bibnamefont {van~den
  Berg}}, \bibinfo {author} {\bibfnamefont {A.}~\bibnamefont {Spinelli}},
  \bibinfo {author} {\bibfnamefont {I.~S.}\ \bibnamefont {Eliens}}, \bibinfo
  {author} {\bibfnamefont {B.}~\bibnamefont {van~den Toorn}}, \bibinfo {author}
  {\bibfnamefont {B.}~\bibnamefont {Bryant}}, \bibinfo {author} {\bibfnamefont
  {J.-S.}\ \bibnamefont {Caux}}, \ and\ \bibinfo {author} {\bibfnamefont
  {A.~F.}\ \bibnamefont {Otte}},\ }\href@noop {} {\bibfield  {journal}
  {\bibinfo  {journal} {Nature Physics}\ } (\bibinfo {year}
  {2016})}\BibitemShut {NoStop}%
\bibitem [{\citenamefont {Vindigni}\ \emph {et~al.}(2006)\citenamefont
  {Vindigni}, \citenamefont {Rettori}, \citenamefont {Pini}, \citenamefont
  {Carbone},\ and\ \citenamefont {Gambardella}}]{Vindigni2006}%
  \BibitemOpen
  \bibfield  {author} {\bibinfo {author} {\bibfnamefont {A.}~\bibnamefont
  {Vindigni}}, \bibinfo {author} {\bibfnamefont {A.}~\bibnamefont {Rettori}},
  \bibinfo {author} {\bibfnamefont {M.}~\bibnamefont {Pini}}, \bibinfo {author}
  {\bibfnamefont {C.}~\bibnamefont {Carbone}}, \ and\ \bibinfo {author}
  {\bibfnamefont {P.}~\bibnamefont {Gambardella}},\ }\href {\doibase
  10.1007/s00339-005-3364-4} {\bibfield  {journal} {\bibinfo  {journal}
  {Applied Physics A}\ }\textbf {\bibinfo {volume} {82}},\ \bibinfo {pages}
  {385} (\bibinfo {year} {2006})}\BibitemShut {NoStop}%
\bibitem [{\citenamefont {Gambardella}\ \emph {et~al.}(2002)\citenamefont
  {Gambardella}, \citenamefont {Dallmeyer}, \citenamefont {Maiti},
  \citenamefont {Malagoli}, \citenamefont {Eberhardt}, \citenamefont {Kern},\
  and\ \citenamefont {Carbone}}]{Gambardella2002}%
  \BibitemOpen
  \bibfield  {author} {\bibinfo {author} {\bibfnamefont {P.}~\bibnamefont
  {Gambardella}}, \bibinfo {author} {\bibfnamefont {A.}~\bibnamefont
  {Dallmeyer}}, \bibinfo {author} {\bibfnamefont {K.}~\bibnamefont {Maiti}},
  \bibinfo {author} {\bibfnamefont {M.~C.}\ \bibnamefont {Malagoli}}, \bibinfo
  {author} {\bibfnamefont {W.}~\bibnamefont {Eberhardt}}, \bibinfo {author}
  {\bibfnamefont {K.}~\bibnamefont {Kern}}, \ and\ \bibinfo {author}
  {\bibfnamefont {C.}~\bibnamefont {Carbone}},\ }\href
  {https://doi.org/10.1038/416301a} {\bibfield  {journal} {\bibinfo  {journal}
  {Nature}\ }\textbf {\bibinfo {volume} {416}},\ \bibinfo {pages} {301}
  (\bibinfo {year} {2002})}\BibitemShut {NoStop}%
\bibitem [{\citenamefont {Landau}\ and\ \citenamefont
  {Lifshitz}(1980)}]{Landau1980}%
  \BibitemOpen
  \bibfield  {author} {\bibinfo {author} {\bibfnamefont {L.}~\bibnamefont
  {Landau}}\ and\ \bibinfo {author} {\bibfnamefont {E.}~\bibnamefont
  {Lifshitz}},\ }\href@noop {} {\emph {\bibinfo {title} {Statistical
  Physics}}},\ \bibinfo {number} {v. 5}\ (\bibinfo  {publisher} {Pergamon
  Press},\ \bibinfo {year} {1980})\BibitemShut {NoStop}%
\bibitem [{\citenamefont {Curilef}\ \emph {et~al.}(2005)\citenamefont
  {Curilef}, \citenamefont {del Pino},\ and\ \citenamefont
  {Orellana}}]{Curilef2005}%
  \BibitemOpen
  \bibfield  {author} {\bibinfo {author} {\bibfnamefont {S.}~\bibnamefont
  {Curilef}}, \bibinfo {author} {\bibfnamefont {L.~A.}\ \bibnamefont {del
  Pino}}, \ and\ \bibinfo {author} {\bibfnamefont {P.}~\bibnamefont
  {Orellana}},\ }\href {\doibase 10.1103/PhysRevB.72.224410} {\bibfield
  {journal} {\bibinfo  {journal} {Phys. Rev. B}\ }\textbf {\bibinfo {volume}
  {72}},\ \bibinfo {pages} {224410} (\bibinfo {year} {2005})}\BibitemShut
  {NoStop}%
\bibitem [{\citenamefont {Li}\ and\ \citenamefont {Liu}(2006)}]{Li2006}%
  \BibitemOpen
  \bibfield  {author} {\bibinfo {author} {\bibfnamefont {Y.}~\bibnamefont
  {Li}}\ and\ \bibinfo {author} {\bibfnamefont {B.-G.}\ \bibnamefont {Liu}},\
  }\href {\doibase 10.1103/PhysRevB.73.174418} {\bibfield  {journal} {\bibinfo
  {journal} {Phys. Rev. B}\ }\textbf {\bibinfo {volume} {73}},\ \bibinfo
  {pages} {174418} (\bibinfo {year} {2006})}\BibitemShut {NoStop}%
\bibitem [{\citenamefont {Kolesnikov}\ \emph {et~al.}(2015)\citenamefont
  {Kolesnikov}, \citenamefont {Tsysar},\ and\ \citenamefont
  {Saletsky}}]{Kolesnikov2015}%
  \BibitemOpen
  \bibfield  {author} {\bibinfo {author} {\bibfnamefont {S.~V.}\ \bibnamefont
  {Kolesnikov}}, \bibinfo {author} {\bibfnamefont {K.~M.}\ \bibnamefont
  {Tsysar}}, \ and\ \bibinfo {author} {\bibfnamefont {A.~M.}\ \bibnamefont
  {Saletsky}},\ }\href {\doibase 10.1134/S1063783415080120} {\bibfield
  {journal} {\bibinfo  {journal} {Physics of the Solid State}\ }\textbf
  {\bibinfo {volume} {57}},\ \bibinfo {pages} {1513} (\bibinfo {year}
  {2015})}\BibitemShut {NoStop}%
\bibitem [{\citenamefont {Onsager}(1944)}]{Onsager1944}%
  \BibitemOpen
  \bibfield  {author} {\bibinfo {author} {\bibfnamefont {L.}~\bibnamefont
  {Onsager}},\ }\href {\doibase 10.1103/PhysRev.65.117} {\bibfield  {journal}
  {\bibinfo  {journal} {Phys. Rev.}\ }\textbf {\bibinfo {volume} {65}},\
  \bibinfo {pages} {117} (\bibinfo {year} {1944})}\BibitemShut {NoStop}%
\bibitem [{\citenamefont {Noble}\ \emph {et~al.}(2015)\citenamefont {Noble},
  \citenamefont {Machta},\ and\ \citenamefont {Hastings}}]{Noble2015}%
  \BibitemOpen
  \bibfield  {author} {\bibinfo {author} {\bibfnamefont {A.~E.}\ \bibnamefont
  {Noble}}, \bibinfo {author} {\bibfnamefont {J.}~\bibnamefont {Machta}}, \
  and\ \bibinfo {author} {\bibfnamefont {A.}~\bibnamefont {Hastings}},\ }\href
  {https://doi.org/10.1038/ncomms7664} {\bibfield  {journal} {\bibinfo
  {journal} {Nature Communications}\ }\textbf {\bibinfo {volume} {6}},\
  \bibinfo {pages} {6664} (\bibinfo {year} {2015})}\BibitemShut {NoStop}%
\bibitem [{\citenamefont {Aguilera}\ and\ \citenamefont
  {Bedia}(2018)}]{Aguilera2018}%
  \BibitemOpen
  \bibfield  {author} {\bibinfo {author} {\bibfnamefont {M.}~\bibnamefont
  {Aguilera}}\ and\ \bibinfo {author} {\bibfnamefont {M.~G.}\ \bibnamefont
  {Bedia}},\ }\href {\doibase 10.3389/fnbot.2018.00055} {\bibfield  {journal}
  {\bibinfo  {journal} {Frontiers in Neurorobotics}\ }\textbf {\bibinfo
  {volume} {12}},\ \bibinfo {pages} {55} (\bibinfo {year} {2018})}\BibitemShut
  {NoStop}%
\bibitem [{\citenamefont {Hoferer}\ \emph {et~al.}(2019)\citenamefont
  {Hoferer}, \citenamefont {Bonfanti}, \citenamefont {Taloni}, \citenamefont
  {La~Porta},\ and\ \citenamefont {Zapperi}}]{Hoferer2019}%
  \BibitemOpen
  \bibfield  {author} {\bibinfo {author} {\bibfnamefont {M.}~\bibnamefont
  {Hoferer}}, \bibinfo {author} {\bibfnamefont {S.}~\bibnamefont {Bonfanti}},
  \bibinfo {author} {\bibfnamefont {A.}~\bibnamefont {Taloni}}, \bibinfo
  {author} {\bibfnamefont {C.~A.~M.}\ \bibnamefont {La~Porta}}, \ and\ \bibinfo
  {author} {\bibfnamefont {S.}~\bibnamefont {Zapperi}},\ }\href {\doibase
  10.1103/PhysRevE.100.042410} {\bibfield  {journal} {\bibinfo  {journal}
  {Phys. Rev. E}\ }\textbf {\bibinfo {volume} {100}},\ \bibinfo {pages}
  {042410} (\bibinfo {year} {2019})}\BibitemShut {NoStop}%
\bibitem [{\citenamefont {Spaide}(2018)}]{Spaide2018}%
  \BibitemOpen
  \bibfield  {author} {\bibinfo {author} {\bibfnamefont {R.~F.}\ \bibnamefont
  {Spaide}},\ }\href {\doibase 10.1097/IAE.0000000000001517} {\bibfield
  {journal} {\bibinfo  {journal} {RETINA}\ }\textbf {\bibinfo {volume} {38}},\
  \bibinfo {pages} {79} (\bibinfo {year} {2018})}\BibitemShut {NoStop}%
\bibitem [{\citenamefont {Shin}\ and\ \citenamefont
  {Kolomeisky}(2019)}]{Shin2019}%
  \BibitemOpen
  \bibfield  {author} {\bibinfo {author} {\bibfnamefont {J.}~\bibnamefont
  {Shin}}\ and\ \bibinfo {author} {\bibfnamefont {A.~B.}\ \bibnamefont
  {Kolomeisky}},\ }\href {\doibase 10.1063/1.5123988} {\bibfield  {journal}
  {\bibinfo  {journal} {The Journal of Chemical Physics}\ }\textbf {\bibinfo
  {volume} {151}},\ \bibinfo {pages} {125101} (\bibinfo {year} {2019})},\
  \Eprint {http://arxiv.org/abs/https://doi.org/10.1063/1.5123988}
  {https://doi.org/10.1063/1.5123988} \BibitemShut {NoStop}%
\bibitem [{\citenamefont {Kadanoff}(1966)}]{Kadanoff1966}%
  \BibitemOpen
  \bibfield  {author} {\bibinfo {author} {\bibfnamefont {L.~P.}\ \bibnamefont
  {Kadanoff}},\ }\href {\doibase 10.1103/PhysicsPhysiqueFizika.2.263}
  {\bibfield  {journal} {\bibinfo  {journal} {Physics Physique Fizika}\
  }\textbf {\bibinfo {volume} {2}},\ \bibinfo {pages} {263} (\bibinfo {year}
  {1966})}\BibitemShut {NoStop}%
\bibitem [{\citenamefont {Widom}(1965)}]{Widom1965}%
  \BibitemOpen
  \bibfield  {author} {\bibinfo {author} {\bibfnamefont {B.}~\bibnamefont
  {Widom}},\ }\href@noop {} {\bibfield  {journal} {\bibinfo  {journal} {The
  Journal of Chemical Physics}\ }\textbf {\bibinfo {volume} {43}},\ \bibinfo
  {pages} {3898} (\bibinfo {year} {1965})}\BibitemShut {NoStop}%
\bibitem [{\citenamefont {Baxter}(1982)}]{Baxter1982}%
  \BibitemOpen
  \bibfield  {author} {\bibinfo {author} {\bibfnamefont {R.~J.}\ \bibnamefont
  {Baxter}},\ }\href@noop {} {\emph {\bibinfo {title} {Exactly solved models in
  statistical mechanics}}}\ (\bibinfo  {publisher} {Academic Press},\ \bibinfo
  {year} {1982})\BibitemShut {NoStop}%
\bibitem [{\citenamefont {Goldenfeld}(1992)}]{Goldenfeld1992}%
  \BibitemOpen
  \bibfield  {author} {\bibinfo {author} {\bibfnamefont {N.}~\bibnamefont
  {Goldenfeld}},\ }\href@noop {} {\emph {\bibinfo {title} {Lectures on phase
  transitions and the renormalization group}}}\ (\bibinfo  {publisher}
  {Addison-Wesley, Advanced Book Program, Reading},\ \bibinfo {year}
  {1992})\BibitemShut {NoStop}%
\bibitem [{\citenamefont {Christensen}\ and\ \citenamefont
  {Moloney}(2005)}]{Christensen2005}%
  \BibitemOpen
  \bibfield  {author} {\bibinfo {author} {\bibfnamefont {K.}~\bibnamefont
  {Christensen}}\ and\ \bibinfo {author} {\bibfnamefont {N.~R.}\ \bibnamefont
  {Moloney}},\ }\href {\doibase 10.1142/p365} {\emph {\bibinfo {title}
  {Complexity and Criticality}}}\ (\bibinfo  {publisher} {Imperial College
  Press},\ \bibinfo {year} {2005})\ pp.\ \bibinfo {pages}
  {140--150}\BibitemShut {NoStop}%
\bibitem [{\citenamefont {Wang}\ and\ \citenamefont {Landau}(2001)}]{Wang2001}%
  \BibitemOpen
  \bibfield  {author} {\bibinfo {author} {\bibfnamefont {F.}~\bibnamefont
  {Wang}}\ and\ \bibinfo {author} {\bibfnamefont {D.~P.}\ \bibnamefont
  {Landau}},\ }\href {\doibase 10.1103/PhysRevE.64.056101} {\bibfield
  {journal} {\bibinfo  {journal} {Phys. Rev. E}\ }\textbf {\bibinfo {volume}
  {64}},\ \bibinfo {pages} {056101} (\bibinfo {year} {2001})}\BibitemShut
  {NoStop}%
\bibitem [{\citenamefont {Caparica}\ and\ \citenamefont
  {Cunha{-}Netto}(2012)}]{Caparica2012}%
  \BibitemOpen
  \bibfield  {author} {\bibinfo {author} {\bibfnamefont {A.~A.}\ \bibnamefont
  {Caparica}}\ and\ \bibinfo {author} {\bibfnamefont {A.~G.}\ \bibnamefont
  {Cunha{-}Netto}},\ }\href@noop {} {\bibfield  {journal} {\bibinfo  {journal}
  {Phys. Rev. E}\ }\textbf {\bibinfo {volume} {85}},\ \bibinfo {pages} {046702}
  (\bibinfo {year} {2012})}\BibitemShut {NoStop}%
\bibitem [{\citenamefont {Caparica}(2014)}]{Caparica2014}%
  \BibitemOpen
  \bibfield  {author} {\bibinfo {author} {\bibfnamefont {A.~A.}\ \bibnamefont
  {Caparica}},\ }\href {\doibase 10.1103/PhysRevE.89.043301} {\bibfield
  {journal} {\bibinfo  {journal} {Phys. Rev. E}\ }\textbf {\bibinfo {volume}
  {89}},\ \bibinfo {pages} {043301} (\bibinfo {year} {2014})}\BibitemShut
  {NoStop}%
\bibitem [{\citenamefont {Ferreira}\ \emph {et~al.}(2018)\citenamefont
  {Ferreira}, \citenamefont {Jorge}, \citenamefont {Leão},\ and\ \citenamefont
  {Caparica}}]{Ferreira2018}%
  \BibitemOpen
  \bibfield  {author} {\bibinfo {author} {\bibfnamefont {L.}~\bibnamefont
  {Ferreira}}, \bibinfo {author} {\bibfnamefont {L.}~\bibnamefont {Jorge}},
  \bibinfo {author} {\bibfnamefont {S.}~\bibnamefont {Leão}}, \ and\ \bibinfo
  {author} {\bibfnamefont {A.}~\bibnamefont {Caparica}},\ }\href {\doibase
  https://doi.org/10.1016/j.jcp.2018.01.003} {\bibfield  {journal} {\bibinfo
  {journal} {Journal of Computational Physics}\ }\textbf {\bibinfo {volume}
  {358}},\ \bibinfo {pages} {130 } (\bibinfo {year} {2018})}\BibitemShut
  {NoStop}%
\bibitem [{\citenamefont {Post}\ \emph {et~al.}(2018)\citenamefont {Post},
  \citenamefont {McLeod}, \citenamefont {Hepting}, \citenamefont {Bluschke},
  \citenamefont {Wang}, \citenamefont {Cristiani}, \citenamefont {Logvenov},
  \citenamefont {Charnukha}, \citenamefont {Ni}, \citenamefont {Radhakrishnan},
  \citenamefont {Minola}, \citenamefont {Pasupathy}, \citenamefont {Boris},
  \citenamefont {Benckiser}, \citenamefont {Dahmen}, \citenamefont {Carlson},
  \citenamefont {Keimer},\ and\ \citenamefont {Basov}}]{Post2018}%
  \BibitemOpen
  \bibfield  {author} {\bibinfo {author} {\bibfnamefont {K.~W.}\ \bibnamefont
  {Post}}, \bibinfo {author} {\bibfnamefont {A.~S.}\ \bibnamefont {McLeod}},
  \bibinfo {author} {\bibfnamefont {M.}~\bibnamefont {Hepting}}, \bibinfo
  {author} {\bibfnamefont {M.}~\bibnamefont {Bluschke}}, \bibinfo {author}
  {\bibfnamefont {Y.}~\bibnamefont {Wang}}, \bibinfo {author} {\bibfnamefont
  {G.}~\bibnamefont {Cristiani}}, \bibinfo {author} {\bibfnamefont
  {G.}~\bibnamefont {Logvenov}}, \bibinfo {author} {\bibfnamefont
  {A.}~\bibnamefont {Charnukha}}, \bibinfo {author} {\bibfnamefont {G.~X.}\
  \bibnamefont {Ni}}, \bibinfo {author} {\bibfnamefont {P.}~\bibnamefont
  {Radhakrishnan}}, \bibinfo {author} {\bibfnamefont {M.}~\bibnamefont
  {Minola}}, \bibinfo {author} {\bibfnamefont {A.}~\bibnamefont {Pasupathy}},
  \bibinfo {author} {\bibfnamefont {A.~V.}\ \bibnamefont {Boris}}, \bibinfo
  {author} {\bibfnamefont {E.}~\bibnamefont {Benckiser}}, \bibinfo {author}
  {\bibfnamefont {K.~A.}\ \bibnamefont {Dahmen}}, \bibinfo {author}
  {\bibfnamefont {E.~W.}\ \bibnamefont {Carlson}}, \bibinfo {author}
  {\bibfnamefont {B.}~\bibnamefont {Keimer}}, \ and\ \bibinfo {author}
  {\bibfnamefont {D.~N.}\ \bibnamefont {Basov}},\ }\href {\doibase
  10.1038/s41567-018-0201-1} {\bibfield  {journal} {\bibinfo  {journal} {Nature
  Physics}\ } (\bibinfo {year} {2018}),\ 10.1038/s41567-018-0201-1}\BibitemShut
  {NoStop}%
\bibitem [{\citenamefont {{\v{S}}im{\.e}nas}\ \emph {et~al.}(2014)\citenamefont
  {{\v{S}}im{\.e}nas}, \citenamefont {Ibenskas},\ and\ \citenamefont
  {Tornau}}]{Simenas2014}%
  \BibitemOpen
  \bibfield  {author} {\bibinfo {author} {\bibfnamefont {M.}~\bibnamefont
  {{\v{S}}im{\.e}nas}}, \bibinfo {author} {\bibfnamefont {A.}~\bibnamefont
  {Ibenskas}}, \ and\ \bibinfo {author} {\bibfnamefont {E.~E.}\ \bibnamefont
  {Tornau}},\ }\href@noop {} {\bibfield  {journal} {\bibinfo  {journal}
  {Physical Review E}\ }\textbf {\bibinfo {volume} {90}},\ \bibinfo {pages}
  {042124} (\bibinfo {year} {2014})}\BibitemShut {NoStop}%
\bibitem [{\citenamefont {Palmer}\ \emph {et~al.}(2018)\citenamefont {Palmer},
  \citenamefont {Poole}, \citenamefont {Sciortino},\ and\ \citenamefont
  {Debenedetti}}]{Palmer2018}%
  \BibitemOpen
  \bibfield  {author} {\bibinfo {author} {\bibfnamefont {J.}~\bibnamefont
  {Palmer}}, \bibinfo {author} {\bibfnamefont {P.}~\bibnamefont {Poole}},
  \bibinfo {author} {\bibfnamefont {F.}~\bibnamefont {Sciortino}}, \ and\
  \bibinfo {author} {\bibfnamefont {P.~G.}\ \bibnamefont {Debenedetti}},\
  }\href {\doibase 10.1021/acs.chemrev.8b00228} {\bibfield  {journal} {\bibinfo
   {journal} {Chem. Rev.}\ }\textbf {\bibinfo {volume} {118}},\ \bibinfo
  {pages} {9129} (\bibinfo {year} {2018})}\BibitemShut {NoStop}%
\bibitem [{\citenamefont {Sloane}(2011)}]{Sloane2011}%
  \BibitemOpen
  \bibfield  {author} {\bibinfo {author} {\bibfnamefont {N.~J.~A.}\
  \bibnamefont {Sloane}},\ }\href@noop {} {\enquote {\bibinfo {title} {{OEIS}
  {F}oundation {I}nc., {T}he {O}n-{L}ine {E}ncyclopedia of {I}nteger
  {S}equences},}\ }\bibinfo {howpublished} {https://oeis.org/A201461} (\bibinfo
  {year} {2011})\BibitemShut {NoStop}%
\bibitem [{\citenamefont {Hanna}(2006)}]{Hanna2006}%
  \BibitemOpen
  \bibfield  {author} {\bibinfo {author} {\bibfnamefont {P.~D.}\ \bibnamefont
  {Hanna}},\ }\href@noop {} {\enquote {\bibinfo {title} {{OEIS} {F}oundation
  {I}nc., {T}he {O}n-{L}ine {E}ncyclopedia of {I}nteger {S}equences},}\
  }\bibinfo {howpublished} {https://oeis.org/A124428} (\bibinfo {year}
  {2006})\BibitemShut {NoStop}%
\bibitem [{\citenamefont {Renner}(2017)}]{Renner2017}%
  \BibitemOpen
  \bibfield  {author} {\bibinfo {author} {\bibfnamefont {M.}~\bibnamefont
  {Renner}},\ }\href@noop {} {\enquote {\bibinfo {title} {{OEIS} {F}oundation
  {I}nc., {T}he {O}n-{L}ine {E}ncyclopedia of {I}nteger {S}equences},}\
  }\bibinfo {howpublished} {https://oeis.org/A281123} (\bibinfo {year}
  {2017})\BibitemShut {NoStop}%
\bibitem [{\citenamefont {Chen}\ \emph {et~al.}(1993)\citenamefont {Chen},
  \citenamefont {Ferrenberg},\ and\ \citenamefont {Landau}}]{Chen1993}%
  \BibitemOpen
  \bibfield  {author} {\bibinfo {author} {\bibfnamefont {K.}~\bibnamefont
  {Chen}}, \bibinfo {author} {\bibfnamefont {A.~M.}\ \bibnamefont
  {Ferrenberg}}, \ and\ \bibinfo {author} {\bibfnamefont {D.~P.}\ \bibnamefont
  {Landau}},\ }\href {\doibase 10.1103/PhysRevB.48.3249} {\bibfield  {journal}
  {\bibinfo  {journal} {Phys. Rev. B}\ }\textbf {\bibinfo {volume} {48}},\
  \bibinfo {pages} {3249} (\bibinfo {year} {1993})}\BibitemShut {NoStop}%
\bibitem [{\citenamefont {Caparica}\ \emph {et~al.}(2000)\citenamefont
  {Caparica}, \citenamefont {Bunker},\ and\ \citenamefont
  {Landau}}]{Caparica2000a}%
  \BibitemOpen
  \bibfield  {author} {\bibinfo {author} {\bibfnamefont {A.~A.}\ \bibnamefont
  {Caparica}}, \bibinfo {author} {\bibfnamefont {A.}~\bibnamefont {Bunker}}, \
  and\ \bibinfo {author} {\bibfnamefont {D.~P.}\ \bibnamefont {Landau}},\
  }\href {\doibase 10.1103/PhysRevB.62.9458} {\bibfield  {journal} {\bibinfo
  {journal} {Phys. Rev. B}\ }\textbf {\bibinfo {volume} {62}},\ \bibinfo
  {pages} {9458} (\bibinfo {year} {2000})}\BibitemShut {NoStop}%
\bibitem [{\citenamefont {Glumac}\ and\ \citenamefont
  {Uzelac}(1989)}]{Glumac1989}%
  \BibitemOpen
  \bibfield  {author} {\bibinfo {author} {\bibfnamefont {Z.}~\bibnamefont
  {Glumac}}\ and\ \bibinfo {author} {\bibfnamefont {K.}~\bibnamefont
  {Uzelac}},\ }\href {\doibase 10.1088/0305-4470/22/20/020} {\bibfield
  {journal} {\bibinfo  {journal} {Journal of Physics A: Mathematical and
  General}\ }\textbf {\bibinfo {volume} {22}},\ \bibinfo {pages} {4439}
  (\bibinfo {year} {1989})}\BibitemShut {NoStop}%
\bibitem [{\citenamefont {Bayong}\ \emph {et~al.}(1999)\citenamefont {Bayong},
  \citenamefont {Diep},\ and\ \citenamefont {Truong}}]{Bayong1999}%
  \BibitemOpen
  \bibfield  {author} {\bibinfo {author} {\bibfnamefont {E.}~\bibnamefont
  {Bayong}}, \bibinfo {author} {\bibfnamefont {H.~T.}\ \bibnamefont {Diep}}, \
  and\ \bibinfo {author} {\bibfnamefont {T.~T.}\ \bibnamefont {Truong}},\
  }\href {\doibase 10.1063/1.370270} {\bibfield  {journal} {\bibinfo  {journal}
  {Journal of Applied Physics}\ }\textbf {\bibinfo {volume} {85}},\ \bibinfo
  {pages} {6088} (\bibinfo {year} {1999})}\BibitemShut {NoStop}%
\bibitem [{\citenamefont {Shen}\ \emph {et~al.}(1997)\citenamefont {Shen},
  \citenamefont {Skomski}, \citenamefont {Klaua}, \citenamefont {Jenniches},
  \citenamefont {Manoharan},\ and\ \citenamefont {Kirschner}}]{Shen1997}%
  \BibitemOpen
  \bibfield  {author} {\bibinfo {author} {\bibfnamefont {J.}~\bibnamefont
  {Shen}}, \bibinfo {author} {\bibfnamefont {R.}~\bibnamefont {Skomski}},
  \bibinfo {author} {\bibfnamefont {M.}~\bibnamefont {Klaua}}, \bibinfo
  {author} {\bibfnamefont {H.}~\bibnamefont {Jenniches}}, \bibinfo {author}
  {\bibfnamefont {S.~S.}\ \bibnamefont {Manoharan}}, \ and\ \bibinfo {author}
  {\bibfnamefont {J.}~\bibnamefont {Kirschner}},\ }\href
  {https://link.aps.org/doi/10.1103/PhysRevB.56.2340} {\bibfield  {journal}
  {\bibinfo  {journal} {Phys. Rev. B}\ }\textbf {\bibinfo {volume} {56}},\
  \bibinfo {pages} {2340} (\bibinfo {year} {1997})}\BibitemShut {NoStop}%
\bibitem [{\citenamefont {Wolf}(2000)}]{Wolf2000}%
  \BibitemOpen
  \bibfield  {author} {\bibinfo {author} {\bibfnamefont {W.}~\bibnamefont
  {Wolf}},\ }\href@noop {} {\bibfield  {journal} {\bibinfo  {journal}
  {Brazilian Journal of Physics}\ }\textbf {\bibinfo {volume} {30}},\ \bibinfo
  {pages} {794} (\bibinfo {year} {2000})}\BibitemShut {NoStop}%
\bibitem [{\citenamefont {Leedahl}\ \emph {et~al.}(2019)\citenamefont
  {Leedahl}, \citenamefont {Sundermann}, \citenamefont {Amorese}, \citenamefont
  {Severing}, \citenamefont {Gretarsson}, \citenamefont {Zhang}, \citenamefont
  {Komarek}, \citenamefont {Maignan}, \citenamefont {Haverkort},\ and\
  \citenamefont {Tjeng}}]{Leedahl2019}%
  \BibitemOpen
  \bibfield  {author} {\bibinfo {author} {\bibfnamefont {B.}~\bibnamefont
  {Leedahl}}, \bibinfo {author} {\bibfnamefont {M.}~\bibnamefont {Sundermann}},
  \bibinfo {author} {\bibfnamefont {A.}~\bibnamefont {Amorese}}, \bibinfo
  {author} {\bibfnamefont {A.}~\bibnamefont {Severing}}, \bibinfo {author}
  {\bibfnamefont {H.}~\bibnamefont {Gretarsson}}, \bibinfo {author}
  {\bibfnamefont {L.}~\bibnamefont {Zhang}}, \bibinfo {author} {\bibfnamefont
  {A.~C.}\ \bibnamefont {Komarek}}, \bibinfo {author} {\bibfnamefont
  {A.}~\bibnamefont {Maignan}}, \bibinfo {author} {\bibfnamefont {M.~W.}\
  \bibnamefont {Haverkort}}, \ and\ \bibinfo {author} {\bibfnamefont {L.~H.}\
  \bibnamefont {Tjeng}},\ }\href {https://doi.org/10.1038/s41467-019-13273-4}
  {\bibfield  {journal} {\bibinfo  {journal} {Nature Communications}\ }\textbf
  {\bibinfo {volume} {10}},\ \bibinfo {pages} {5447} (\bibinfo {year}
  {2019})}\BibitemShut {NoStop}%
\bibitem [{\citenamefont {Šimėnas}\ \emph {et~al.}(2015)\citenamefont
  {Šimėnas}, \citenamefont {Ibenskas},\ and\ \citenamefont
  {Tornau}}]{Simenas2015}%
  \BibitemOpen
  \bibfield  {author} {\bibinfo {author} {\bibfnamefont {M.}~\bibnamefont
  {Šimėnas}}, \bibinfo {author} {\bibfnamefont {A.}~\bibnamefont {Ibenskas}},
  \ and\ \bibinfo {author} {\bibfnamefont {E.~E.}\ \bibnamefont {Tornau}},\
  }\href {\doibase 10.1080/01411594.2014.983509} {\bibfield  {journal}
  {\bibinfo  {journal} {Phase Transitions}\ }\textbf {\bibinfo {volume} {88}},\
  \bibinfo {pages} {833} (\bibinfo {year} {2015})},\ \Eprint
  {http://arxiv.org/abs/https://doi.org/10.1080/01411594.2014.983509}
  {https://doi.org/10.1080/01411594.2014.983509} \BibitemShut {NoStop}%
\end{thebibliography}
%
\end{document}